\documentstyle[hx37_latex]{article}
\input{psfig}
\begin{document}
\title{ A GLOBAL PERSPECTIVE ON STAR FORMATION }
\author{ S. MICHAEL FALL }
\address{ Space Telescope Science Institute \\ 3700 San Martin Drive \\
Baltimore, MD 21218, USA }

\def\E{{\cal E}}
\def\lesssim{\mathrel{\hbox to 0pt{\lower 3.5pt\hbox{$\mathchar"218$}\hss}
      \raise 1.5pt\hbox{$\mathchar"13C$}}}
\def\gtrsim{\mathrel{\hbox to 0pt{\lower 3.5pt\hbox{$\mathchar"218$}\hss}
      \raise 1.5pt\hbox{$\mathchar"13E$}}}

\maketitle\abstracts{
We outline a method to infer the global history of star formation in
galaxies with input only from absorption-line observations of quasars.
The application of the method to existing data leads to the conclusion 
that most stars formed at relatively low redshifts ($z\lesssim2$). 
We combine the global rate of star formation with stellar population 
synthesis models to compute the mean comoving emissivity and mean 
intensity of background radiation from far-UV to far-IR wavelengths. 
These predictions are consistent with all the available measurements 
and observational limits, including recent results from {\it HST} and 
{\it COBE}.
}

\section{Overview}

This article concerns the evolution of, and relations between, various 
large-scale average properties of the population of galaxies as a whole.
It is often convenient to express these ``global'' properties as mean
comoving densities and to normalize them to the present closure density.
We are particularly interested in the comoving densities of stars, gas, 
metals, and dust within galaxies, which we denote respectively by 
$\Omega_s$, $\Omega_g$, $\Omega_m$, and $\Omega_d$. 
The last three of these are meant to refer to the interstellar media (ISM)
of galaxies, exclusive of the intergalactic medium (IGM), although in
practice such a distinction may only be approximate.
As defined here, $\Omega_m$ includes metals in both the gas and solid 
(i.e., dust) phases of the ISM.
It is usually more informative to reexpress $\Omega_m$ and $\Omega_d$ in 
terms of the mean metallicity and mean dust-to-gas ratio, $Z\equiv\Omega_m
/\Omega_g$ and $D/G\equiv\Omega_d/\Omega_g$.
It is clear that all of these properties are related in the sense that, as 
new stars form, $\Omega_s$ will increase, while, in most cases, $\Omega_g$
will decrease, and $Z$ and $D/G$ will increase.
One of our goals is to quantify such relations through the equations of 
``cosmic chemical evolution''. 

Until recently, there were no emission-based estimates of the global rate 
of star formation $\dot\Omega_s$ at $z\gtrsim0.3$.
The reason for this is that samples of galaxies selected by emission 
become progressively incomplete and include only brighter objects at 
higher redshifts.
In contrast, samples of galaxies selected by absorption against 
background quasars do not suffer from this bias. 
Such observations are exquisitely sensitive to small column densities 
of absorbing or scattering particles.
In principle at least, they enable us to estimate $\Omega_g$, $\Omega_m$, 
and even $\Omega_d$ as functions of redshift. 
From these and the equations of cosmic chemical evolution, we can then 
infer the global rate of star formation $\dot\Omega_s$. 
It is amusing to note that this idealistic program does not require the
detection of a single stellar photon!
Furthermore, if we are confident (or foolish) enough, we can combine 
our estimates of $\dot\Omega_s$ with stellar population synthesis models
to compute the mean comoving emissivity $\E_\nu$ and the mean intensity 
of background radiation $J_\nu$.
One might then claim to have predicted the ``emission history'' of the 
universe from its ``absorption history''. 
This article describes a first attempt by Yichuan Pei, St\'ephane Charlot, 
and the author to carry out such a program; a complete account of our work
is given in references 1 and 2. 
Some related material can be found in references 3--6.

\section{Absorption-Line Systems}

Before proceeding, it is worth recalling some facts about the statistics 
of absorption-line systems.
Let $f(N_x,z)$ be the column density distribution of particles of any
type $x$ that absorb or scatter light.
These might, for example, be hydrogen atoms ($x=$~HI), metal ions ($x=m$), 
or dust grains ($x=d$). 
By definition, $H_0 (1+z)^3 |dt/dz| f(N_x,z)dN_xdz$ is the mean number 
of absorption-line systems with column densities of $x$ between $N_x$ and 
$N_x+dN_x$ and redshifts between $z$ and $z+dz$ along the lines of sight 
to randomly selected background quasars.
These lines of sight are very narrow (much less than a parsec across) and
pierce the absorption-line systems at random angles and impact parameters.
One can show that the mean comoving density of $x$ is given by 
$$
\Omega_x(z) = {8 \pi G m_x \over 3 c H_0} \int_0^\infty dN_x f(N_x,z) N_x,
\eqno(1)
$$
where $m_x$ is the mass of a single particle (atom, ion, or grain).
Equation~(1) plays a central role in this subject.
It enables us to estimate the mean comoving densities of many quantities
of interest without knowing anything about the structure of the 
absorption-line systems.
In particular, we do not need to know their sizes or shapes, whether they
are smooth or clumpy, and so forth.
A corollary of equation~(1) is that the global metallicity, $Z\equiv
\Omega_m/\Omega_g$, is given simply by an average over the metallicities 
of individual absorption-line systems weighted by their gas column 
densities.

The absorption-line systems of most interest in the present context are 
the damped Ly$\alpha$ (DLA) systems. 
It is widely believed that they trace the ISM of galaxies and protogalaxies 
and are the principal sites of star formation in the universe.
There are excellent reasons to adopt this as a working hypothesis. 
First, the DLA systems have, by definition, $N_{\rm HI}\gtrsim
10^{20}$~cm$^{-2}$, which is just below an apparent threshold for the 
onset of star formation~\cite{ken89}.
Second, the DLA systems contain at least 80\% of the HI in the universe 
and appear to be mostly neutral~\cite{lan95}.
The other absorption-line systems, those with $N_{\rm HI}\lesssim
10^{20}$~cm$^{-2}$, probably contain more gas in total than the DLA
systems, but this must be diffuse and mostly ionized.
In the following, we regard non-DLA systems as belonging to the IGM,
even though some of them might actually be located in the outer halos
of galaxies.
This distinction -- between the mostly-neutral ISM, where stars form, and
the mostly-ionized IGM, where they do not -- is clearly valid at the 
present epoch.
Thus, the DLA systems are often referred to as DLA galaxies.
It will be interesting to see exactly which types of galaxies they 
represent, but as we have already emphasized, this issue does not 
affect any of the global properties derived from equation~(1).

The sample of known DLA galaxies now includes about 80 
objects~\cite{wol95}.
They are distributed over a wide range in redshift, $0 \lesssim z 
\lesssim 4$, although, as a consequence of selection effects, most of 
them are confined to the narrower range $2 \lesssim z \lesssim 3$.
From observations of DLA galaxies in various subsets of this sample 
and comparisons with present-day galaxies, the following trends have 
emerged.
The mean comoving density of HI decreases by almost an order of magnitude,
from $\Omega_{\rm HI}\approx(1-2)\times10^{-3}h^{-1}$ at $z\approx3$ to 
$\Omega_{\rm HI}\approx2\times10^{-4}h^{-1}$ at $z=0$ [with $h\equiv
H_0/(100$~km~s$^{-1}$~Mpc$^{-1}$)]~\cite{lan95,wol95,sto96}.
It is possible that $\Omega_{\rm HI}$ increases between $z\approx4$ and
$z\approx3$, but the evidence for this is weak~\cite{sto96}.
The mean metallicity increases by about an order of magnitude, from 
$Z\approx0.1Z_\odot$ or slightly less at $z\approx2$ to $Z\approx 
Z_\odot$ at $z=0$~\cite{pet94,pet96}.
The mean dust-to-gas ratio increases by a similar factor, while the mean
dust-to-metals ratio remains roughly constant at about the present value 
in the local ISM~\cite{pet94,pet96,pei91}.
These results are entirely consistent with the recent Keck observations 
by Lu et al~\cite{pet96,lue96,kul96}. 
The abundances of H$_2$ and CO appear to be much lower at $z\gtrsim2$ than 
at $z=0$~\cite{lev92}.
As a consequence of the relatively small samples involved, most of the 
numbers quoted here are uncertain by factors of 1.5 or more.

\section{Cosmic Chemical Evolution}

The global properties defined above are governed by a set of coupled 
equations, which are sometimes referred to as the equations of cosmic 
chemical evolution. 
In the approximation of instantaneous recycling (and $Z\ll1$), they take
the form
$$
{d\over{dt}}(\Omega_g+\Omega_s) = \dot\Omega_f, \eqno(2) 
$$
$$
{d\over{dt}}(Z\Omega_g) + (Z-y){d\over{dt}}\Omega_s = Z_f\dot\Omega_f,
\eqno(3)
$$ 
where $y$ is the IMF-averaged yield. 
Equations~(2) and (3) are strictly valid only when all galaxies evolve 
in the same way;
otherwise, they should be regarded as approximations.
The ``source'' terms on the right-hand sides of the equations allow for 
the exchange of material between the ISM of galaxies and the IGM; they 
represent the inflow or outflow of gas with metallicity $Z_f$ at a rate 
$\dot\Omega_f$.
To illustrate a range of possibilities, we consider three types of 
evolution: a closed-box model ($\dot\Omega_f=0$), a model with inflow of 
metal-free gas ($\dot\Omega_f=+\nu\dot\Omega_s$, $Z_f=0$), and a model 
with outflow of metal-enriched gas ($\dot\Omega_f=-\nu\dot\Omega_s$, 
$Z_f=Z$). 
Our inflow and outflow models are direct analogs of the standard models 
of chemical evolution in the disk and spheroid components of the Milky 
Way~\cite{lar72,har76}. 
We fix the yield $y$ in each model by requiring $Z=Z_\odot$ at $z=0$. 
Then the only adjustable parameters are the ``initial'' comoving density 
of gas in galaxies $\Omega_{g\infty}$ (in practice, the value of 
$\Omega_g$ at $z\gtrsim4$) and the relative inflow or outflow rate $\nu$. 

To complete the specification of the models, we make two other
approximations, both motivated by the observations summarized in the 
previous section.
(1)~We neglect any ionized or molecular gas in the ISM of galaxies and 
set $\Omega_g=1.3\Omega_{\rm HI}$ (to account for He).
(2)~We assume that just over half of the metals in the ISM are depleted
onto dust grains and set $D/G=0.6Z$.
The models are designed to reproduce (as input) the observed decrease
in the mean comoving density of HI between $z\approx3$ and $z=0$.
The only subtlety here is that the observed values of $\Omega_{\rm HI}$ 
tend to underestimate the true values as a consequence of the obscuration 
of quasars by dust in foreground galaxies~\cite{fal93}.
We make a self-consistent correction for this bias in the models by 
linking the obscuration of quasars to the chemical enrichment of galaxies.
It is worth noting that, while this correction has a substantial effect 
on $\Omega_{\rm HI}$, especially at $z\sim1$, it does not entail large 
numbers of ``missing'' quasars (only $\sim$20\% at $z=2$ and $\sim$40\% 
at $z=4$).
The models reproduce (as output) the observed increase in the mean 
metallicity between $z\approx2$ and $z=0$ without any fine tuning
of the parameters $\Omega_{g\infty}$ and $\nu$.
The reason for this is that most of the star formation and hence most
of the metal production occur at $z\lesssim2$.

Figure~1 shows the evolution of the comoving rate of metal production
$\dot\rho_z$ in the models. 
This is given by $\dot\rho_z=y\psi$, with $\psi=(1-R)^{-1}\dot\rho_s$
and $\dot\rho_s=(3H_0^2/8\pi G)\dot\Omega_s$, where $R\approx0.3$ is the
returned fraction.
The predicted rates have maxima at $1\lesssim z \lesssim 2$ and decline 
rapidly at lower redshifts.
Figure~1 also shows estimates of, and lower limits on, $\dot\rho_z$ from 
recent ground-based surveys and the Hubble Deep 
Field~\cite{gal95,lil96,ste96,mad96}.
These are proxies for global H$\alpha$ and UV emissivities based on the 
close correspondence between UV emission and metal production in massive 
stars~\cite{cow88,mad96}.
Evidently, the predicted and observed rates are in broad qualitative,
and even some quantitative, agreement (given the uncertainties in both).
This is remarkable because the models were constructed only with 
absorption-line systems in mind, not the emissivities represented 
in Figure~1.
We have also combined our chemical evolution models with stellar 
population synthesis models to compute directly the mean comoving
emissivity $\E_\nu$ at wavelengths from $10^{-1}\mu$m to $10^3\mu$m 
and, by an integration over redshift, the corresponding mean intensity 
of background radiation $J_\nu$. 
These calculations include a self-consistent treatment of the 
absorption and reradiation of starlight by the dust within galaxies.
The same models shown in Figure~1 also predict a far-IR/sub-mm 
background in nice agreement with a tentative detection based on 
{\it COBE} data~\cite{pug96}.

\begin{figure}
\vskip-70pt
\psfig{figure=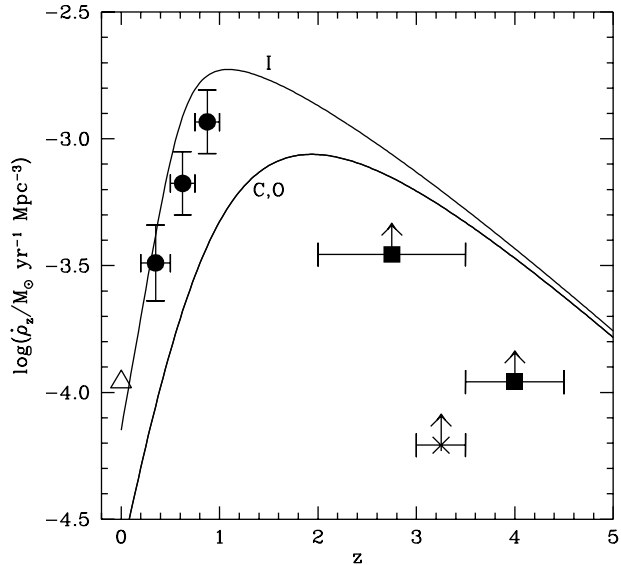,width=4in}
\vskip-10pt
\caption{
Comoving rate of metal production $\dot\rho_z$ as a function of redshift 
$z$ (for $h=0.5$, $q_0=0.5$, and $\Lambda=0$). 
The curves are from the closed-box (C), inflow (I), and outflow (O) models 
with $\Omega_{g\infty}=4\times10^{-3}h^{-1}$ and $\nu=0.5$ (see Figure~1 
of reference~1). 
The data points and lower limits represent global H$\alpha$ and UV 
emissivities from ground-based and {\it HST} surveys (see Figure~9 of 
reference~22).
}
\end{figure}

\end{document}